\documentstyle[aps,pra,psfig,epsf,epsfig,twocolumn,floats]{revtex}

\begin{document}

\draft
\title{Matter-wave entanglement and teleportation by molecular dissociation 
and collisions}
\author{T. Opatrn\'{y}$^{1}$  and  G. Kurizki
 }
\address{
Department of Chemical Physics, Weizmann Institute of Science,
761~00 Rehovot, Israel \\
$^{1}$ 
Theor. Phys. Institut, F. Schiller Universit\"{a}t,
Max-Wien-Platz 1, D-07743 Jena, Germany \\
and  Theor.
Phys. Dept., Palack\'{y} University, Svobody 26, 77146 Olomouc, Czech Republic 
}
\date{\today}

\maketitle

\begin{abstract}
We propose dissociation of cold diatomic molecules as a
source of atom pairs with highly correlated (entangled) positions and momenta,
approximating the original quantum state introduced by Einstein, Podolsky and 
Rosen (EPR) [Phys. Rev. {\bf 47,} 777 (1935)]. Wavepacket teleportation 
is shown to be achievable by its collision with one of the EPR correlated 
atoms and manipulation of the other atom in the pair.
\end{abstract}

The fundamentally profound notion of quantum teleportation is the prescription
of how to map, in a {\em one to one fashion, any quantum state\/}  of system A
onto that of a {\em distant\/} system B: one must measure the pertinent
observables of  A, then manipulate their counterparts in B  according to the
communicated results of the measurements on A
\cite{Bennett,Vaidman,Zeilinger,Boschi,BK98,Furusawa,Kuzmich}. 
Teleportation has thus
far been explored for photon polarizations 
\cite{Bennett,Vaidman,Zeilinger,Boschi},
optical field quadratures \cite{BK98,Furusawa} and multi-atom spin components 
\cite{Kuzmich}. Here we pose the basic question: How to teleport the {\em
translational quantum states (wavepackets) of  material bodies\/}   over
sizeable distances? We propose dissociation of cold diatomic molecules as a
source of atom pairs with highly correlated (entangled) positions and momenta,
approximating the original quantum state introduced by Einstein, Podolsky and 
Rosen (EPR) \cite{EPR}. Wavepacket teleportation \cite{Vaidman}
is shown to be achievable by 
its collision with one of the EPR correlated atoms and manipulation of the
other atom in the pair.

\begin{figure}[!t!]
\noindent
\centerline{\epsfig{figure=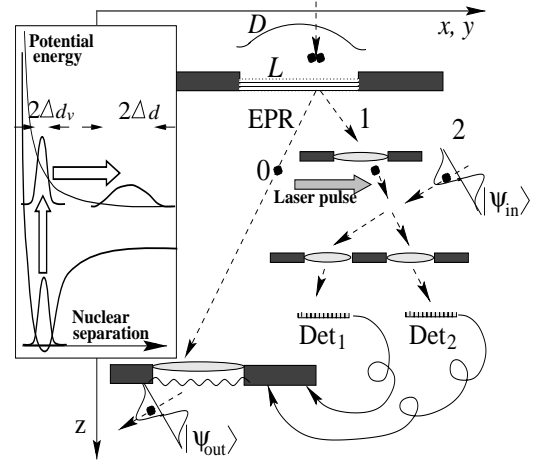,width=0.8\linewidth}}
\vspace{2ex}
\caption{
EPR entanglement and teleportation of atomic wavepackets. A cold diatom with
COM spread $D$ dissociates into translationally entangled atoms or ions 0 and
1. Particle 1 is focussed and laser-deflected to collide with particle 2. Their
post-collision distance and momentum-sum are measured by detectors 1 and 2
and determine the position and momentum shifts of particle 0,
whose translational state $|\psi_{\rm out}\rangle$ then approximately
reproduces  $|\psi_{\rm in}\rangle$ of particle 2. Inset: short-pulse
dissociation yields a replica of the internuclear wavepacket on the
dissociative potential surface, whose gradient causes the wavepacket to spread.
}
\label{eprsch3}
\end{figure}

Consider a cold beam of ionized molecules that are moving with high, constant 
velocity $v_{z}$ along the $z$-axis (which thus {\em plays the role of time})
to a region where they are dissociated by means of an  electromagnetic pulse.
Let the size $L$ of the dissociation region along the perpendicular $x$-axis
be defined by means of an aperture (Fig.~\ref{eprsch3}). Here the
molecule splits and its two constituents  start receding.

In order to be coherent within the dissociation region $L$, the translational
{\em center-of-mass\/} (COM) state of the molecule along the  $x$-axis should
be close to the minimum uncertainty state (MUS) of its position and momentum,
described by a Gaussian exp$[-P_x^2/2\Delta P_{x}^{2}]$, where $P_x$ is the
$x$-component of the COM momentum and $\Delta P_x$ its spread. Such a state
can be prepared by translationally cooling the molecules to the ground state
of a trapping potential,   then ionizing  and accelerating them to the
required speed $v_{z}$ (prior to dissociation). The required temperature is
typically $T$ $\approx$ $\hbar^{2}/(M k_{B} D^{2})$, where $M$ is the molecular
mass, $k_{B}$ is the Boltzmann constant and $D$, the COM wavepacket size, is
chosen to be $D$ $\lesssim$ $L$. A size $D$ $\approx$  300 nm would require $T$
$\approx$ 3~$\mu$K for H$_{2}^{+}$ and  $T$ $\approx$ 0.4~$\mu$K for 
Li$^-_{2}$. Such temperatures are achievable by Raman photoassociation of 
atomic pairs in Bose condensates \cite{Heinzen}.

We would like the two-atom translational state obtained by molecular
dissociation to closely resemble the {\em original} EPR state \cite{EPR},
whose realization has not been investigated thus far. The perfect EPR state of
particles 0 and 1 is  described by the wavefunction 
$\Psi(x_{0},x_{1})=\delta (x_{0}-x_{1})$: 
the positions and momenta of the two particles along
$x$ are completely uncertain, yet perfectly correlated (entangled):  $x_{0} =
x_{1}$, $p_{x0} = -p_{x1}$. The resemblance of the dissociated state to the
perfect EPR state depends on the extent to which the variances of the
correlated atomic variables are {\em below} the Heisenberg uncertainty limit,
so that 
\begin{eqnarray} 
 \Delta x_{01} \Delta P_{x} \ll \hbar ,
 \label{eqheis} 
\end{eqnarray} 
$\hat x_{01}$ $=$ $\hat x_0$ $-$  $\hat x_1$ and
$\hat P_{x}$ $=$ $\hat p_{x_0}$ $+$  $\hat p_{x_1}$ being respectively the 
internuclear separation and COM momentum operators. Inequality (\ref{eqheis})
is possible since the two variances {\em do not\/} pertain to canonically
conjugate variables. The diatomic molecule prior to dissociation is
describable by an internuclear wavepacket whose spatial size is $\Delta d_v$
(typically 0.1 nm),
determined by its vibrational state. Dissociation by a short laser pulse can
yield an {\em almost exact replica\/} of the initial  internuclear wavepacket
on a dissociative electronic potential surface of the molecule (Fig.
\ref{eprsch3}-inset). The much broader, cold COM wavepacket keeps during
dissociation the molecular MUS spread (Fig.~\ref{eprsch3}) $\Delta P_{x}$
$\approx$ $\hbar/D$. As the internuclear wavepacket {\em becomes
nearly free\/} shortly after dissociation, it may still resemble an EPR
state. Its proximity to the perfect EPR state at $t$ $=$ $0$, chosen to
signify the {\em completion\/} of dissociation, can be measured by the  
parameter $s$ that is inversely proportional to the left-hand side of
(\ref{eqheis}) \begin{eqnarray}
  s \approx  D/\Delta d.
  \label{squeezing}
\end{eqnarray}
Here the width $\Delta d$ $=$ $\Delta x_{01}(0)$ $>$ $\Delta d_v$ is
determined by the dynamical spreading of the internuclear wavepacket, caused
by the gradient of the dissociative potential surface. The ratio in
(\ref{squeezing}) should be sought to satisfy $s\gg 1$, in accordance with
inequality (\ref{eqheis}), the perfect EPR state having $s$ $\to$ $\infty$.
Even for the realistic values $\Delta d \sim 1$\ nm $\gg \Delta d_v$, and
$D$ $\agt$ 300\ nm, this parameter is remarkably large: $s$   $\approx$ $300$.

\begin{figure}[!t!]
\noindent
\centerline{\epsfig{figure=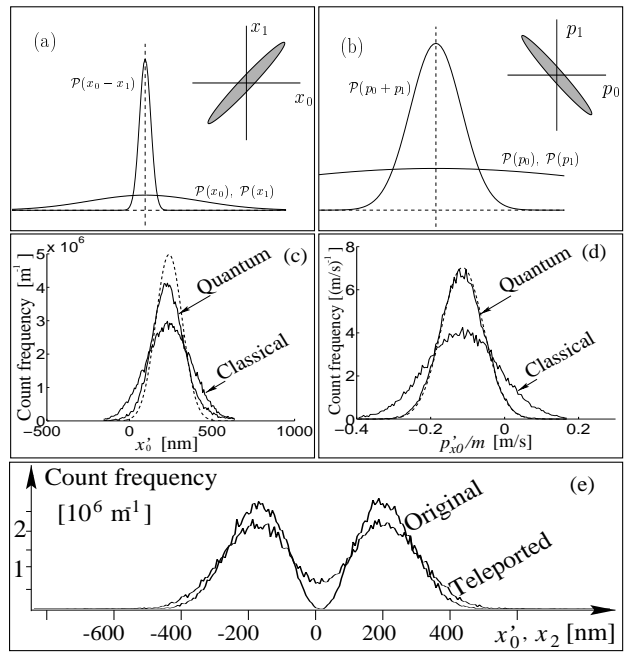,width=0.95\linewidth}}
\vspace{2ex}
\caption{
Probability distributions of the separation ${\cal P}(x_0-x_1)$ (a) and
momentum-sum  ${\cal P}(p_0+p_1)$ (b) of the EPR pair, as compared with their
respective single-particle counterparts ${\cal P}(x_{0(1)})$ and
${\cal P}(p_{0(1)})$. 
(c)-(e): Our prediction for the results of $5\times 10^{4}$ teleportation
events with lithium ions with the parameters as in the text. (c) position, and
(d) momentum distribution of a Gaussian wavepacket, broken line: ideal input
distributions, full lines: results of classical and quantum 
teleportations (by following classical trajectories of an
ensemble obeying the quantum Wigner $x-p$ distribution). 
(e): Prediction of the teleported position distribution 
of a two-peak interference pattern (by applying the
calculated teleportation-induced noise to the input distribution). 
} 
\label{f-correls}
\end{figure}

There is a noteworthy analogy between the two-particle EPR state and
that of entangled two-mode light from parametric downconversion (PDC) 
\cite{Reid,Ou}: $s$ is analogous to the exponential of the ``squeezing 
parameter'' of such light, which represents the ratio of the standard 
deviation of the ``stretched'' field quadrature to that of its ``squeezed'' 
counterpart (one quadrature corresponding, e.g., to the sum and the other 
to the difference of the  two fields, analogously to our $\hat{x}_{01}$ and 
$\hbar /\hat{P}_{x}$, respectively). 
In current optical experiments \cite{Ou} $s$ $\lesssim$4.

Outside the dissociation region (typically, for separations beyond a few nm)
the receding particles evolve freely, with diminishing position correlation.
Yet the two-particle wavefunction remains inseparable, having the form $\Psi
(x_0,x_1)$ $=$ $\psi(x_{01})\psi_{\rm COM}(x_0 + x_1)$, i.e., a product of the
internuclear wavefunction and its COM counterpart. The spread of the coordinate
difference $x_{01}$ then  grows with time as $\Delta x_{01}^2(t)$  $=$ $\Delta
d^2$ $+$ $\Delta v_{01}^2t^2$, $\Delta v_{01}$ being the spread of their
velocity difference  acquired during dissociation (here we assume homonuclear
diatomic molecules, for simplicity).  Yet the growth of $\Delta x_{01}(t)$ {\em
can be compensated\/} by a suitable lens, which {\em ideally} ``time-reverses''
$\Delta x_{01}$, i.e.,  projects  it back to its initial ``spot'' of size
$\Delta x_{01}(0)$ $=$ $\Delta d$. In practice, such compensation is limited by
the resolution of the focussing lens (see below).

We suggest that the translationally-entangled particles 0 and 1 can be used 
to demonstrate the {\em hitherto unobserved\/} original prediction of EPR 
\cite{EPR}, concerning the nonlocality of quantum correlations in a 
free-propagating two-atom state satisfying inequality (\ref{eqheis}): 
detecting the momentum or position of particle 1 would make the corresponding 
variable of particle 0 assume a nearly well-defined value (with fluctuations  
$\hbar/ D$ or $\Delta d$, respectively). By contrast, there will be large 
fluctuations if we measure {\em uncorrelated\/} variables (Fig. \ref{f-correls}
(a), (b)): the momentum of particle 1 and the position of particle 0, or vice 
versa. We note that the preparation of EPR states of {\em internal atomic 
observables\/} (unlike the present external or translational observables) by 
diatomic dissociation was discussed in \cite{Kurizki,Kurizki2} (for atomic 
excitations or pseudospin states), and in \cite{Fry} (for atomic spin states).
An EPR correlation of {\em trapped\/} (rather than free-propagating) atoms is 
proposed in \cite{Parkins}.


The major challenge is to teleport the quantum state $|\psi_{in}\rangle$ of
the {\em transversal\/} motion of particle 2 (along the $x$ axis) to particle
0 (Fig.~\ref{eprsch3}). To this end, particle 2 collides with one member of
the EPR pair -- particle 1 (in a synchronous fashion determined by laser 
pulses---see below), after which both particles 1 and 2 are detected.
Essentially, the collision of particles 1 and 2 allows us to project their
pre-collisional {\em joint state\/}   onto the basis of EPR-correlated  states
(specified below) and detect the result of the projection.   The results of
the post-collision detection are used to control the evolution of the 
``readout'' particle 0, the other member of the EPR pair. 
In the optimal case, the resulting
translational state $|\psi _{out}\rangle$ of particle 0 would closely imitate
the input $|\psi_{in}\rangle$ of particle 2. We may assume that
$|\psi_{in}\rangle$  is prepared by  diffraction on a double-slit screen or 
grating \cite{Mlynek}.  If measured directly, the beam of particle 2 would
exhibit a characteristic diffraction pattern. If the teleportation scheme is
applied instead, then the input beam is destroyed but nearly the  same
diffraction pattern can be observed in the transformed output beam of particle
0. We note that the collision region in Fig.~\ref{eprsch3}  plays an {\em
analogous role to the beam-splitter\/} which mixes a field quadrature of the
teleported optical state with the field of one of the entangled PDC modes in
the teleportation scheme of \cite{BK98}.
 
A natural choice for the post-collision correlated (EPR-state) basis is the
set of states associated with a sharp (well-defined) sum of the colliding
particles' momenta $\hat p_{x1}$ $+$ $\hat p_{x2}$  and sharp difference of
their $x$ coordinates, $\hat x_1$ $-$ $\hat x_2$. At the time instant $\tau$,
corresponding to the particles' nearest approach in the absence of interaction
(see below), the operators $\hat x_{-}$ and $\hat p_{+}$, given by
\begin{eqnarray}
 \hat x_{-} & \equiv &  \hat x_1 (\tau)- \hat x_2(\tau) 
 \label{dverce1} \\
 \hat p_{+} &  \equiv & 
 \hat p_{x1}(\tau) + \hat p_{x2}(\tau)
 \label{dverce} 
\end{eqnarray}
are measured, with values $x_-$, $p_+$ and precision 
$\Delta x_{\rm meas}$ and $\Delta p_{\rm meas}$, respectively 
(which are discussed below).
 
The last step required for the teleportation of the state of particle 2 is 
the shift of the $x$-coordinate and momentum of the ``readout'' particle 0, 
according to the inferred values of $x_{-}$ and $p_{+}$ as  
$\hat x_0$ $\to$ $\hat x'_0$ $=$ $\hat x_0 -  x_{-}$, and
$\hat p_{x0}$ $\to$ $\hat p'_{x0}$ $=$ $\hat p_{x0} +  p_{+}$. 
Let the precision of the
position and momentum shifters be $\Delta x_{\rm shift}$
and $\Delta p_{\rm shift}$, respectively.  
{F}rom the correlations of $\hat x_0$ and $\hat x_1$,
or $\hat p_{x0}$ and $\hat p_{x1}$ discussed above, 
we find that (at the relevant
times) the shifted variables satisfy
\begin{eqnarray} 
 \label{x0x2}
 \hat x'_0 &=& \hat  x_2  \pm \Delta x_T, \\ 
 \label{p0p2}
 \hat p'_{x0} &=& \hat  p_{x2}  \pm \Delta p_T,  
\end{eqnarray} 
namely, they approximately {\em reproduce the teleported variables of 
particle\/} 2, to the accuracy given by  $\Delta x_T$ and $\Delta p_T$
where
\begin{eqnarray} 
 \label{xspread} 
 \Delta x_T^2 = \Delta d^{2} + \Delta x_{\rm meas}^2
 + \Delta x_{\rm shift}^2  , 
 \\
 \Delta p_T^2 =  \Delta P_x^2 + \Delta p_{\rm meas}^2 
 + \Delta p_{\rm shift}^2.
 \label{pspread}  
\end{eqnarray} 
Equations (\ref{x0x2}), (\ref{p0p2}) imply that the Wigner $x-p$ distribution
$W(x'_0,p'_{x0})$ of the readout particle 0 is given by that of the input
particle 2, $W_{in}(x_2,p_{x2})$, coarse-grained  by a smoothing function whose 
width is determined by  Eqs.~(\ref{xspread}), (\ref{pspread}). 

The teleportation has non-classical properties provided:
(i)   $\Delta x_T$ and $\Delta p_T$  satisfy
\begin{eqnarray}
 \Delta x_T \Delta p_T < \hbar ; 
 \label{telepcond}
\end{eqnarray}
(ii) they are finer than the scales of
the input wavepacket variation in position and momentum, respectively 
[Fig.~\ref{f-correls}(c)-(e)]. 
If the fluctuations produced by the measurements and shifts are
negligible ($\Delta x_{\rm meas}$, $\Delta x_{\rm shift}$
$\ll$ $\Delta d$, and $\Delta p_{\rm meas}$, $\Delta p_{\rm shift}$
$\ll$ $\Delta P_x$), then the left-hand side of Eq. (\ref{telepcond})
is approximately $\Delta x_T \Delta p_T$ $\approx \hbar /s$.
For teleported Gaussian wavepackets, the maximum attainable fidelity, i.e.,
the overlap of the input state with the teleported output, can be shown to be 
$F_{\rm max}$ $=$ $(1+\Delta x_T \Delta p_T/\hbar)^{-1}$. 
As shown in \cite{BK98}, the maximal {\it classical} teleportation fidelity is
1/2, so that larger fidelity, as in the quantum-optical teleportation 
experiment \cite{Furusawa} where $F= 0.58$, attests to quantum correlations.
[``Classical teleportation'' is based on simultaneous (though noisy)
measurements of the non-commuting quantities $\hat x_2$ and $\hat p_{x2}$ and
preparing a corresponding MUS wavepacket of particle 0.]

\begin{figure}[!t!]
\noindent
\centerline{\epsfig{figure=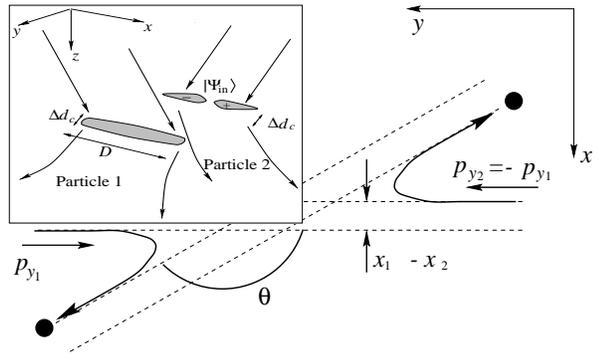,width=0.9\linewidth}}
\vspace{2ex}
\caption{
Collision scheme of particles 1 and 2. Their deflection angle $\theta$
corresponds to the collision distance $x_1-x_2$. Their momentum sum is
conserved. Inset: structure of wavepackets 1 and 2, marked by width $\Delta
d_c$ along the $y$ and $z$ axes and carrying quantum information along the
$x$ axis. }   
\label{f-collis} 
\end{figure}

We proceed to discuss the experimentally relevant aspects of the envisaged
teleportation. The dissociated pairs of fragments (particles) 0 and 1 can be
emitted in all directions,  but the dissociating field polarization can impose
directionality  \cite{Larsen}. A significant fraction of the fragment pairs
enter the aperture of the lens focussing them onto the region where the
collision of particles 1 and 2 takes place. The colliding particles are
assumed to be prepared in {\em uncorrelated\/} wavepackets propagating with
the same classical velocity along $z$, $v_{z1}$ $=$ $v_{z2}$, and opposite
classical momenta along $y$ (Fig.~\ref{f-collis}), $m_1 v_{y1}$ $=$
$-m_2 v_{y2}$, such that $v_{z1,2}\gg|v_{y1,2}|\gg \Delta v_{x1,2}$.
Their focussing (and laser-pulse deflection) is such that 
{\em in the absence of interaction}
the wavepackets would cross each other in the $xy$ plane, where their size
would be  smallest (``contractive'' wavepackets \cite{sto94}). Let us take
both particles 1 and 2 to be equally charged. Charged-particle focussing (by
electrostatic and  magnetic lenses used in high-resolution microscopy) would
render the spread  (``spot size'') of the colliding particles' positions as
sharp as it was at  the relevant times (thus achieving ``time-reversal''): for
particle 1 this time  corresponds to the completion of
dissociation, whereas for the  teleported particle 2 it is the time of the
state preparation (e.g., by a double slit). The arrival of the two particles
in the $xy$ collision plane at $\tau$ must be {\it synchronized by laser 
pulses}, causing the fast  switching-on of their deflection towards the 
collision region in the $x-y$ plane, provided the values of $z_1$, $z_2$, 
$v_{z_1}=v_{z_2}$ are appropriate. 
Synchronization  accuracy of $\sim$10 ps and $v_z$
$\sim$ $10^3$ m/s would bring the particles well within the Coulomb collision
range (see below). The $y-z$ extent of the two wavepackets must be $\Delta
d_{c}$ $\ll$ $D$, so that they overlap in the absence of Coulomb repulsion
(Fig.~\ref{f-collis}--inset). Among the dissociated atom pairs entering the 
input aperture $L$, only those participating in the {\em synchronized\/} 
collision events are counted as pairs that contribute to successful
teleportation.

The success of teleportation hinges upon our ability to discriminate, when
detecting the post-collision states, between correlated EPR states with {\em
different values of the relevant parameters,\/} namely, the parameter
$x_{-}$ and the momentum sum $p_{+}$ in Eqs.~(\ref{dverce1}) and 
(\ref{dverce}). We can infer  the parameter   $x_{-}$ by measuring the relative
deflection angle of the colliding particles. A quantitative estimation can be
made for repulsive Coulomb interaction of two particles with mass $m$ and
charge $q$. Their relative deflection angle is given by (see, e.g.,
\cite{Bransden}) $\theta$ $=$ $\pi$ $-$  $2$ $\tan ^{-1}$ $\left[
(x_{1}-x_{2})/R_{\rm col} \right]$, where $R_{\rm col}$ $=$ $m
q^{2}/(4\pi\epsilon_{0} p_{y}^{2})$ is the characteristic collision range. To
ensure maximal sensitivity of the deflection angle  to $x_{-}$ it is useful to
choose the experimental parameters such that $\Delta d_{c}$ $\ll$ $R_{\rm col}$
$\sim$ $D \sim$ ${\rm max}|x_-|$.  From the collisional analysis it follows
that the precision of inferring $x_{-}$ is mostly affected by the fluctuation
of its conjugate momentum component $\Delta (p_{x2}-p_{x1})$ $\sim$
$\hbar/\Delta d_v$. The corresponding resolution of $x_{-}$, obtained from
that of $\theta$, is $\Delta x_{\rm meas}$  
$\approx$ $\pi \epsilon_0 \hbar D^2 p_y/(2 q^2 m \Delta d_v)$,
provided that $p_y$ $\gtrsim$ $\sqrt{m q^2/(4\pi \epsilon_0 D)}$. 
In particular, for lithium ions with  
velocities $v_{y} \approx 300$ m/s,  $R_{\rm col} \approx 220$nm,  
the resolution of position difference measurements is
$\Delta x_{\rm meas}$ $\approx 15$nm. 
The momentum-sum $p_{+}$  is measurable by
the Doppler shift of Raman transitions induced by two counterpropagating laser
fields, with  precision better than 1\ mm/s \cite{Kasevich}. The dominant
contribution to its resolution, $\Delta p_{\rm meas}$, then stems from the 
fluctuation $\Delta P_x$,
which is, for the given parameters, $\Delta P_x/m$ $\approx$ 30\ mm/s. 

In the last stage of teleportation, particle 0  must be ``kicked''
(momentum-shifted)  and position-shifted
 by external fields. The spread incurred by these shifts is
incorporated into Eqs.~(\ref{xspread}) and (\ref{pspread}), 
and constitutes one of the factors in the
anticipated resolution used in the teleportation simulations of
Fig.~\ref{f-correls}(c). Yet the precision of ion optics is currently so high 
that the dominant contributions in $\Delta x_T$ and $\Delta p_T$ 
stem from $\Delta x_{\rm meas}$ and $\Delta p_{\rm meas}$, respectively. 
We then get for the teleportation of lithium ion wavepackets,
under the conditions specified above, 
the estimation $\Delta x_T \Delta p_T/\hbar$ $\approx 0.08$ $\ll 1$. 
Such spreads would allow for much higher teleportation fidelities of continuous
variables than those presently achievable in quantum optical experiments.

The foregoing analysis shows that, despite the challenging nature of the 
proposed ideas, their experimental implementation does not pose 
insurmountable (rather than technical) difficulties. 
The teleportation of matter wavepackets by molecular dissociation
and collisions is a novel concept, combining elements of molecular dynamics 
and ion (atom) optics, and having quantum optical analogues. 
Its realization is a challenging
but viable goal to pursue, en route to quantum information exchange between 
complex material objects. 
 
{\em Acknowledgments:} The support of US-Israel BSF and DFG is
acknowledged. G.~K. holds the G.~W.~Dunne Professorial Chair.




\begin{thebibliography}{99} 
 

\bibitem{Bennett} 
C. Bennett  et al., 
 Phys. Rev. Lett. {\bf 70,} 1895 (1993). 
 
\bibitem{Vaidman}
L. Vaidman,
 Phys. Rev. A {\bf 49,} 1473 (1994).
 
 \bibitem{Zeilinger} 
D. Bouwmeester  et al., 
 Nature {\bf 390,} 575  (1997).

\bibitem{Boschi} 
D. Boschi  et al., 
 Phys. Rev. Lett. {\bf 80,} 1121  (1998). 
 
\bibitem{BK98} 
S. L. Braunstein and  H. J.  Kimble,
 Phys. Rev. Lett. {\bf 80,} 869  (1998).

\bibitem{Furusawa} 
A. Furusawa   et al., 
 Science  {\bf 282,} 706 (1998). 

\bibitem{Kuzmich}
 A. Kuzmich and  E. S.  Polzik,
Atomic quantum state teleportation and swapping. 
arXiv: quant-ph/0003015 (2000).
  
\bibitem{EPR} 
 A. Einstein,  B. Podolsky, and  N.  Rosen, 
 Phys. Rev. {\bf 47,} 777 (1935). 
 
\bibitem{Heinzen}
R. Wynar  et al., 
 Science {\bf 287,} 1016 (2000). 
 
\bibitem{Reid} 
 M. D. Reid and  P. D. Drummond, 
 Phys. Rev. Lett. {\bf 60,} 2731  (1988). 

\bibitem{Ou} 
 Z. Y.  Ou et al., 
 Phys. Rev. Lett. {\bf 68,} 3663 (1992). 

\bibitem{Kurizki}
 G. Kurizki and A. Ben-Reuven, 
 Phys. Rev. A {\bf 32,} 2560  (1985). 

\bibitem{Kurizki2}
G. Kurizki,   
Phys. Rev. A {\bf 43,} R2599  (1991).

\bibitem{Fry}
E. S. Fry, T. Walther,   and S. F. Li,  
 Phys. Rev. A {\bf 52,} 4381 (1995). 
 
\bibitem{Parkins} 
A. S. Parkins and H. J.  Kimble,  
Phys. Rev. A {\bf 61,} 2104 (2000).
 
\bibitem{Mlynek} 
S. Kunze, K.  Dieckmann,  and G. Rempe,  
 Phys. Rev. Lett. {\bf 78,} 2038  (1997). 

\bibitem{Larsen}
 J. J. Larsen  et al., 
 Phys. Rev. Let. {\bf 83,} 1123  (1999). 
 
\bibitem{sto94} 
 P. Storey et al., 
 Phys. Rev. A {\bf 49,} 2322 (1994). 
 
\bibitem{Bransden} 
B. H. Bransden,  Atomic Collision Theory 2nd ed. (Benjamin/Cummings,  
Reading, Massachusetts, 1983). 

\bibitem{Kasevich}
M. Kasevich,  et al. 
 Phys. Rev. Lett. {\bf 66,} 2297  (1991).
 
 
\end{thebibliography}
\end{document}